\documentclass[%
 reprint,
superscriptaddress,
onecolumn,
nofootinbib,
amsmath,
amssymb,
aps,
pra,
longbibliography
]{revtex4-2}

\usepackage{placeins}
\usepackage{xspace}
\usepackage{appendix}
\usepackage{graphicx}
\usepackage{dcolumn}
\usepackage{bm}
\usepackage{booktabs}
\usepackage{hyperref}
\usepackage{xcolor}
\hypersetup{
    colorlinks,
    linkcolor={blue!70!black},
    citecolor={blue!70!black},
    urlcolor={blue!70!black}
}

\newcommand*{\wn}{\ensuremath{\text{cm}^{-1}}\xspace}

\newcommand*{\muB}{\ensuremath{\mu_B}\xspace}

\newcommand*{\mysim}{\ensuremath{\sim \!}}

\newcommand*{\XSigma}{\ensuremath{\tilde{X}\,{}^2  \Sigma^+}\xspace}
\newcommand*{\APiHalf}{\ensuremath{\tilde{A}\,{}^2 \Pi_{1/2}}\xspace}

\newcommand*{\ketX}[1]{\ensuremath{\lvert \XSigma, #1 \rangle}\xspace}
\newcommand*{\ketA}[1]{\ensuremath{\lvert \APiHalf, #1 \rangle}\xspace}

\begin{document}

\title{Zeeman-Sisyphus Deceleration for Heavy Molecules with Perturbed Excited-State Structure}
\author{Hiromitsu Sawaoka$^*$}
\author{Alexander Frenett$^*$}
\author{Abdullah Nasir}
\author{Tasuku Ono}
\author{Benjamin L. Augenbraun}
\affiliation{Department of Physics, Harvard University, Cambridge, MA 02138, USA}
\affiliation{Harvard-MIT Center for Ultracold Atoms, Cambridge, MA 02138, USA}

\author{Timothy C. Steimle}
\affiliation{School of Molecular Sciences, Arizona State University, Tempe, AZ 85287, USA}

\author{John M. Doyle}
\affiliation{Department of Physics, Harvard University, Cambridge, MA 02138, USA}
\affiliation{Harvard-MIT Center for Ultracold Atoms, Cambridge, MA 02138, USA}

\def\thefootnote{*}\footnotetext{These authors contributed equally to this work.}\def\thefootnote{\arabic{footnote}}

\date{October 18, 2022}

\begin{abstract}
\noindent We demonstrate and characterize Zeeman-Sisyphus (ZS) deceleration of a beam of ytterbium monohydroxide (YbOH). Our method uses a combination of large magnetic fields ($\mysim 2.5$~T) and optical spin-flip transitions to decelerate molecules while scattering only $\sim 10$ photons per molecule. We study the challenges associated with the presence of internal molecular perturbations among the excited electronic states and discuss the methods used to overcome these challenges, including a modified ZS decelerator using microwave and optical transitions.

\end{abstract}

\maketitle

\section{Introduction} 

The rich internal structures of molecules can be utilized to precisely probe for physics beyond the Standard Model (BSM)~\cite{demille2015diatomic,Cairncross2019}. Current and planned experiments search for fundamental symmetry violations~\cite{Hudson2011, ACME2014, ACME2018, Cairncross2017, Norrgard2019}, time-variation of fundamental constants, and dark matter~\cite{Hanneke2020, Kobayashi2019, jansen2014perspective, Kozlov2013Linear,Kozyryev2021Enhanced}. In many cases molecules with heavy atomic constituents are used because heavy nuclei cause relativistic enhancements that provide higher intrinsic sensitivity to fundamental symmetry violating effects originating from BSM particles (e.g. electric dipole moments (EDM) of elementary particles). Recent proposals have also targeted polyatomic molecules, which generically possess nearly degenerate rotational-vibrational states. Such structure leads to high molecular polarization at low electric fields, parity doublets (useful for EDM experiments), as well as high sensitivty to time variation of fundamental constants.~\cite{kozyryev2017PolyEDM, Kozyryev2021Enhanced,Norrgard2020MgNC}. These features, which are helpful for precision measurements, may also lead to perturbations that make it technically challenging to achieve the necessary level of quantum state control.


Taking full advantage of molecules for precision measurements requires cooling them to ultracold ($\lesssim 100~\mu$K) temperatures. Low temperatures suppress broadening mechanisms and allows for trapping and concomitant long interaction times~\cite{Tarbutt2018, McCarron2018}. One approach to creating ultracold molecules is laser cooling, where molecules are first cooled cryogenically (creating a cold beam), then radiatively slowed, and then loaded into a magneto-optical trap (MOT). Additional methods can then be used for cooling and loading into traps, for example an optical dipole trap (ODT). These steps have led to the successful cooling and loading into an ODT of diatomic (SrF, CaF, YO) and triatomic (CaOH) molecules~\cite{norrgard2015sub, truppe2017CaF, Collopy2018, Vilas2022}, and the magnetic trapping of CaF~\cite{Williams2018magtrap}. Radiative slowing is widely recognized as one of the most difficult steps in this process because, under typical conditions, $\sim 10^4$ photons per molecule are required, a constraint that leads to technical challenges due to leakage into dark states. Transverse pluming of the molecular beam due to many photon momentum kicks also leads to loss of molecular flux.

Many polyatomic molecules, including those proposed for next-generation BSM measurements~\cite{kozyryev2017PolyEDM, Augenbraun2021Observation, AugenbraunYbOHSisyphus, mitra2020direct, Yu2021Probing, Norrgard2020MgNC, Aggarwal2018}, can be easily photon cycled hundreds of times---but cycling $10^{4}$ photons is challenging due to these molecules' complex internal structures. Because radiative slowing consumes the most photons in the laser cooling sequence, finding alternative (non-radiative) methods for the slowing step would open the methods of MOT loading and optical trapping to a broader set of molecules. 

One alternative to radiative slowing is a deceleration method using magnetic fields and single optical pumping in a "Zeeman-Sisyphus" (ZS) configuration~\cite{comparat2014molecular, Fitch2016}. This method builds on previously demonstrated magnetic trap loading techniques~\cite{lu2014magnetic}, extended to multiple stages. The first realization of the ZS decelerator leveraged the large ($\mysim 1$--$3$~K) energy shifts experienced by paramagnetic molecules in Tesla-level magnetic fields, as demonstrated on the polyatomic species CaOH. Nearly $10\%$ of a molecular beam was decelerated to velocities sufficiently low for direct MOT loading~\cite{augenbraun2021Zeeman}. The average slowed molecule in that work scattered fewer than 10 photons per molecule, meaning the ``photon budget'' for deceleration to MOT capture velocities was reduced by a factor of $\sim10^3$. Open questions about the method remained, however, such as whether the deceleration method will be equally applicable to heavy molecules, particularly when their excited electronic states are perturbed and/or possess large magnetic $g$-factors. 

In this paper, we use YbOH to study the application of ZS deceleration to heavier molecules with more complex structure. We first review the ZS deceleration scheme, and then present a spectroscopic characterization of the YbOH level structure in magnetic fields up to $\sim$2.5~T. Next, we describe how this structure affects the efficiency of the ZS deceleration. The features discussed are expected to be generic to many heavy molecules of current interest.  Finally, we demonstrate near-optimal optical pumping efficiency and deceleration of YbOH using an abbreviated ZS decelerator, accumulating molecules near 20~m/s. Our results show that ZS deceleration can be extended to heavy polyatomic molecules, but care must be paid to the complex level structure in planning the experiment.

\section{Experimental Setup}
\begin{figure}[tb]
    \centering 
    \includegraphics[width=0.6\columnwidth]{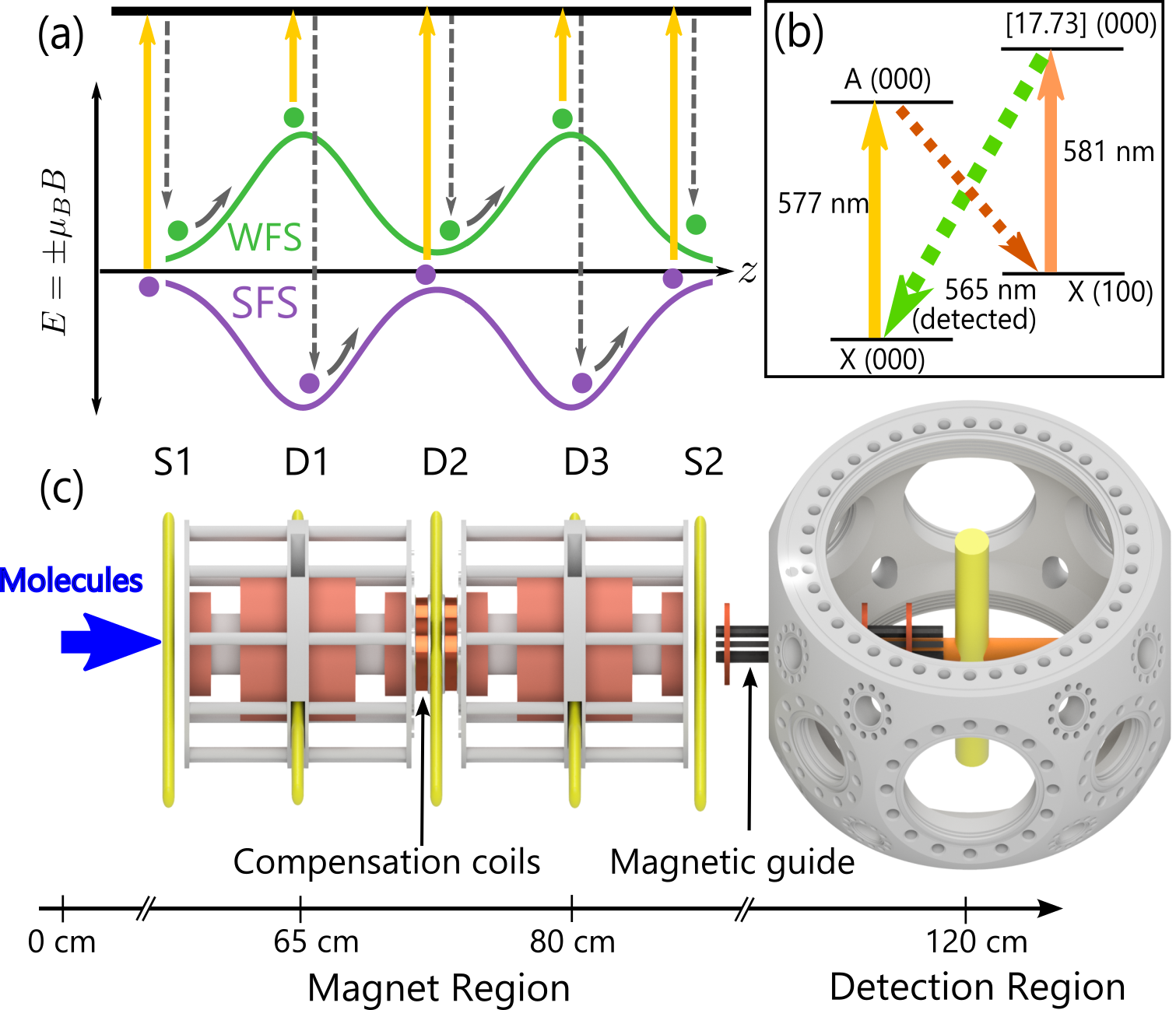}
    \caption{Overview of the ZS deceleration scheme. (a) Simplified representation of the magnetic field tuning for an ideal $^2\,\Pi - ^2\Sigma^+$ electronic transition. (b) Level diagram indicating the background-free detection scheme.  (c) Schematic rendering of the decelerator magnets and detection region.}
    \label{fig:ZSscheme}
\end{figure}

The ZS decelerator (see Fig. \ref{fig:ZSscheme}) has been described in detail previously~\cite{augenbraun2021Zeeman}. In brief, molecules in a WFS state are incident on a region of increasing magnetic field magnitude and decelerate as they proceed into higher magnetic fields. Near the magnetic field maximum, the molecules are optically pumped through an electronically excited state to a SFS state and continue to decelerate as they exit the high-field region. The process can be repeated to remove more energy. In the experiments described here, we use a two-stage decelerator.

In this paper, we denote specific energy levels in the $^2\Sigma^{+}$ ground state using $\ketX{v_1'',N'',M_s''}$, where $v_1''$ is the quanta in the first vibration mode (other vibration modes have 0 quanta for the states we are interested in), $N''$ is the rotational quantum number in Hund's case (b), and $M_s''$ is either strong-field-seeking (SFS) or weak-field-seeking (WFS) at nonzero magnetic fields. The parity of the state is determined by $p'' = (-1)^{N''}$. If not specified, $v_1'' = 0$ and $N'' = 1$ is assumed. In the excited $^2\Pi_{1/2}$ state we use the notation $\ketA{v_1',J',p'}$, where $v_1'$ is the quanta in the first vibration mode, $J'$ is the total electronic angular momentum in Hund's case (a), and $p'$ is the parity. If not specified, $v_1' = 0$, $J' = 1/2$ and $p' = +$ is assumed.

YbOH molecules are produced by laser ablation of a target pressed from a stoichiometric mixture of Yb and Yb(OH)$_3$ powders. The YbOH molecules are produced inside a buffer gas cell held at $\sim$2 K and filled with He buffer gas at a typical atom density of $10^{15}$~cm$^{-3}$. The molecules thermalize to the cell temperature and are hydrodynamically extracted from the cell through a 7~mm diameter aperture~\cite{hutzler2012buffer}. The molecules then enter a second gas collisional cell that is 20~mm long~\cite{hutzler2012buffer,Barry2011} held at $0.9$~K by a pumped $^3$He pot. A 2~mm gap between the first and second cells is used to tune the buffer gas density in the second stage, optimizing the molecular beam's forward velocity and flux. Typical peak forward velocities after the second cell are between 30 and 50~m/s (equivalent kinetic energy $\sim15$~K).

The molecules propagate 55~cm to the decelerator, which consists of two superconducting Helmholtz magnets. Before entering the first magnet, molecules in region S1 are optically pumped into the WFS manifold by a state-preparation laser (7~mm diameter, 100 mW) tuned to the \ketX{SFS}$\rightarrow$\ketA{J'=1/2,p'=+} transition. The molecules then enter the first magnet, decelerating as they gain magnetic potential energy in the increasing field. At the center of the first magnet (region D1, $B \approx 2.4$~T), molecules are optically pumped toward the SFS manifold by a laser beam tuned to the \ketX{WFS}$\rightarrow$\ketA{J'=1/2,p'=+} transition. The laser beam in region D1 contains 400~mW and is cylindrically expanded to waists $\sim$ 20~mm$\times$5~mm, nearly the maximum size compatible with magnet design. Molecules in the SFS manifold continue to decelerate as they exit the first magnet. When the molecules reach region D2, they are driven back to the WFS manifold via the \ketX{SFS}$\rightarrow$\ketA{J'=1/2,p'=+} transition. The laser beam in region D2 has a power of 200~mW and is also expanded to $\sim$ 20~mm$\times$5~mm.  
An extra pair of compensation coils are installed to ensure $B < 500$~G in this region so as to maximize optical pumping efficiency, as explained in section~\ref{sec:OpticalPumping}. The high-field pumping step is repeated in the second magnet as molecules pass through region D3. Immediately after exiting the second magnet, a final spin-flip transition is performed in region S2, putting molecules into a WFS state before they enter a magnetic octupole guide the spans the distance between the decelerator output and the detection region.

Molecules are detected using a (nearly) background-free laser induced fluorescence (LIF) detection scheme. We simultaneously drive the $\APiHalf (000) \leftarrow \XSigma (000)$ (577~nm) and $[17.73](000) \leftarrow \XSigma(000)$ (581~nm) transitions\footnote{The $[17.73]$ state is an electronically excited state present in YbOH that arises from excitations of the Yb($4f$) shell. It behaves mostly like a $\Pi_{1/2}$ state and decays predominantly to the $\XSigma(000)$ and $(100)$ levels.} while detecting fluorescence from $[17.73](000) \rightarrow \XSigma(000)$ at 565~nm.\footnote{A small magnetic shim coil is used to keep the magnitude of the field in the detection region $\lesssim 1$~G.} The 581~nm laser beam can either be sent transverse to the molecular beam, to detect all velocity classes, or can counterpropagate opposite the molecular beam direction, to provide velocity-sensitive detection. In the Doppler-sensitive configuration, the 581~nm beam is chopped at 500 Hz and sequential comparison of time bins with the laser on and off are used to reduce the noise associated with scattered light from the 577 nm beam, which is the dominant source of noise.

\section{Zeeman Spectroscopy}

We performed optical pumping measurements using components of the ZS decelerator to characterize the YbOH Zeeman levels in magnetic fields up to 2.5~T. In these measurements, molecules are first prepared in the WFS manifold in region S1 and enter region D1 in which a (variable) magnetic field is applied. The molecular flux transmitted to the detection region is monitored as a function of the frequency of the optical pumping laser for various magnetic fields.

Because SFS states are filtered out by the octupole guide (see Fig.~\ref{fig:ZSscheme}), resonant transitions that drive population from the WFS manifold to the SFS manifold are detected as a decrease in detected LIF.

\begin{figure}[tb]
    \centering 
    \includegraphics[width=1\columnwidth]{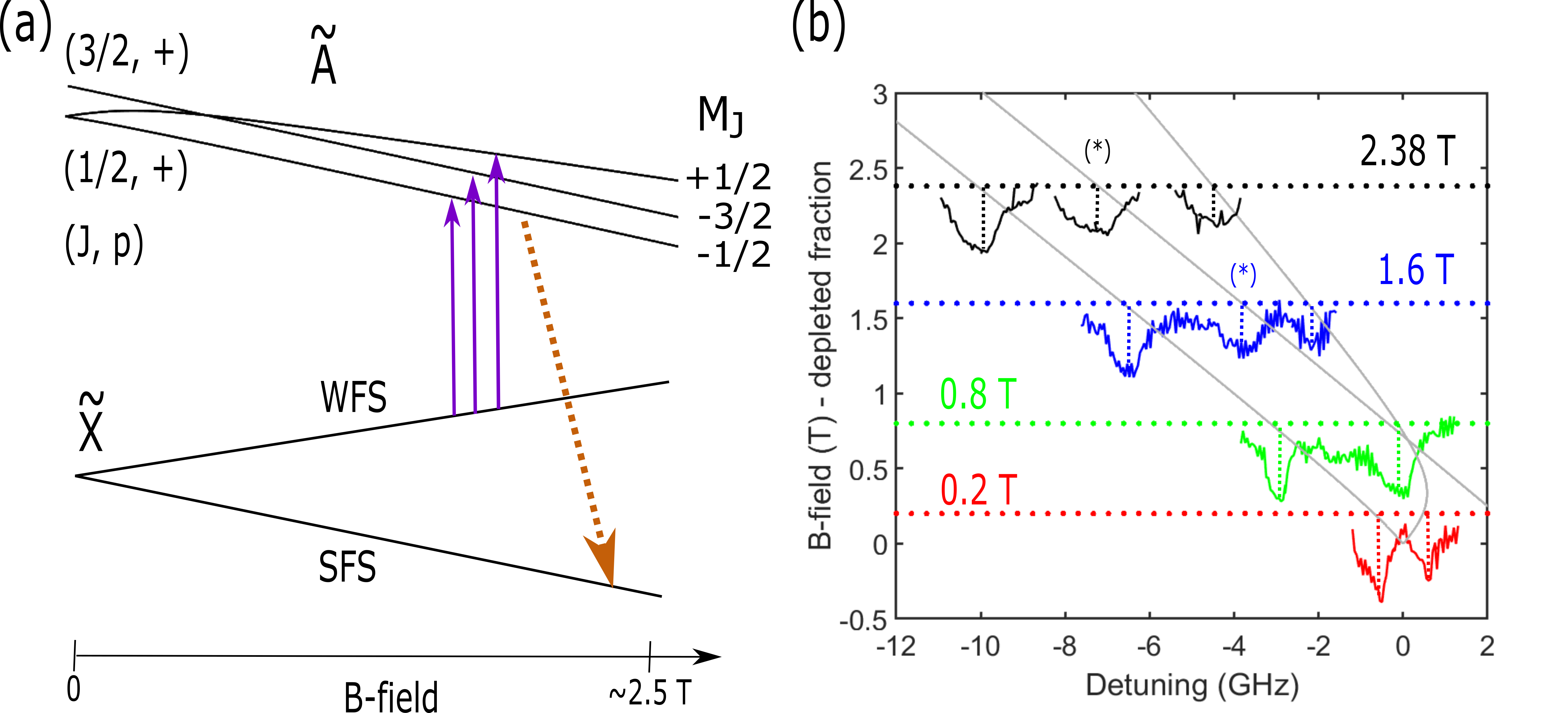}
    \caption{Overview of the \APiHalf(000) Zeeman structure and optical Zeeman spectroscopy. (a) Energy levels and excitation/decay pathways involved in the optical pumping transitions. (b) Magnetic tuning of the low-$J$ \APiHalf(000) energy levels inferred from the optical pumping measurements Data recorded at different magnetic fields are offset vertically for clarity. Lines overlaid on the spectra indicate computed energy levels using optimized fit parameters. The vertical dotted lines are guides to the eye for indicating the center of the depletion features at each magnetic field values. (*) denotes the resonance due to the $\ketA{J'=3/2,p'=+}$ level.}
    \label{fig:AStateZeeman} 
\end{figure}

Figure~\ref{fig:AStateZeeman}(b) shows optical pumping spectra recorded at various magnetic fields between $0.2$~T and $2.4$~T. By subtracting the known ($1\muB$) ground state Zeeman shifts from the raw spectra, we isolate the $\APiHalf$ energy level structure in Figure \ref{fig:AStateZeeman}(b). The $\ketA{J'=1/2,p'=+}$ manifold displays significant Zeeman tuning. Moreover, the nonlinear Zeeman shifts indicate significant rotational mixing with $\ketA{J'=3/2,p'=+}$. The appearance of a third resonance, marked by (*) in Figure~\ref{fig:AStateZeeman}, is due to the $\ketA{J'=3/2,p'=+}$ level, which tunes strongly toward the lower energy $J'=1/2$ states. We fit the observed transitions to the Hamiltonian model of Ref.~\cite{Steimle2019}. Holding all constants other than $g_S$ and $g'_l$ in the \APiHalf state fixed, we determine $g_S = 1.860(9)$ and $g'_l = -0.724(4)$. $g_S$ significantly differs from the bare electron value, although this is not unexpected given perturbations between the \APiHalf state and nearby Yb$^{+}(4f^{13}6s^{2})$OH$^{-}$ states (see supplemental material for additional details). The measured $g'_l$ value is close to that predicted by a Curl-type relationship~\cite{Curl1965}, which predicts $g'_l = -0.865$. The fitting method and obtained values are discussed in detail in supplemental material.

The ground state can also experience level crossings that lead to rotational mixing. At fields greater than $\mysim 0.01$~T, the two Zeeman manifolds of the $\ketX{N''=1}$ level tune linearly with a slope of $1\muB$. Near 0.5~T, the $\ketX{N''=1,\text{SFS}}$ manifold crosses the $\ketX{N''=0,\text{WFS}}$ manifold and near 1~T, the $\ketX{N''=1,\text{WFS}}$ manifold crosses the $\ketX{N''=2,\text{SFS}}$ manifold. However, these crossings do not affect the $\ketX{N''=1}$ manifolds because the neighboring rotational states have opposite parity. On the other hand, near 2.5~T, the $\ketX{N''=1,\text{WFS}}$ manifold crosses the $\ketX{N''=3,\text{SFS}}$ manifold and these levels \textit{can} mix due to dipolar hyperfine terms. For YbOH, 2.5~T is the practical maximum field that can be used for the peak field of a deceleration stage. In general, the crossing between the $\ketX{N''=1,\text{WFS}}$ manifold and the $\ketX{N''=3,\text{SFS}}$ manifold happens at a magnetic field of $5B/\muB $, where $B$ is the rotational constant in units of energy.

\section{Optical Pumping Efficiency} \label{sec:OpticalPumping}

Efficient spin-flip transitions are of central importance to ZS deceleration. For this reason, the initial proposals~\cite{Fitch2016, comparat2014molecular} and demonstration~\cite{augenbraun2021Zeeman} for the ZS decelerator focused on molecules whose lowest electronic excited states are nearly ideal $ ^2\Pi_{1/2}$ states. The large spin-orbit coupling, small $g$-factor, and the presence of rotationally-closed cycling transitions with $^2\Sigma$ ground states makes such electronic states prime targets for ZS deceleration. 

In the $^2\Sigma$ ground state, the electron spin is uncoupled from the molecular axis, giving rise to WFS and SFS states with well defined values of $M_S$, the projection of the electron spin along the external field. Pumping population between these states involves flipping the projection of the electron spin. The large spin-orbit coupling of a $^2\Pi_{1/2}$ state facilitates driving these spin flips because eigenstates in the excited state are associated with a linear combination of $M_S$ values.

By driving transitions between $\ketX{N''=1}$ and $\ketA{J'=1/2,p'=+}$ state, parity and angular momentum selection rules ensure that molecules always decay back to the $N''=1$ manifold. The rotational closure of a cycling transition is crucial to minimize loss during the repeated optical pumping steps. This situation, which has been used in our previous ZS deceleration of CaOH~\cite{augenbraun2021Zeeman} and molecular laser cooling experiments~\cite{Tarbutt2018}, relies on the $\ketA{J'=1/2,p'=+}$ states to be well separated from other levels, ensuring that there is negligible Zeeman mixing with higher-$J$ states and also that transitions that decay to other rotational states are not accidentally resonant in an optical pumping region.

The primary challenge in extending ZS deceleration to YbOH molecules comes from the complex structure of its \APiHalf excited state. This state has an effective $g$-factor, $g_J = g'_l/3 \approx -0.25$, that is over an order of magnitude larger than that of CaOH~\cite{Steimle2019, augenbraun2021Zeeman}. Furthermore, the large $\Lambda$-type doubling in YbOH places the $\ketA{J'=1/2,p'=+}$ and $\ketA{J'=3/2,p'=+}$ levels of the \APiHalf state within about 0.1~\wn of one another. Thus, the \APiHalf states of YbOH experiences both large, non-linear Zeeman shifts and significant state mixing among the \APiHalf rotational levels. The ZS deceleration scheme must be re-examined in light of these features, both of which may reduce the optical pumping efficiency.

Even in low fields where the mixing among these rotational levels is not significant, care must be paid to avoid accidental resonances to lossy transitions through the $\ketA{J'=3/2,p'=+}$ states. In particular, at $\sim 1000$ G, the field used in region D2 for our previous ZS deceleration of CaOH \cite{augenbraun2021Zeeman}, a transition from the WFS manifold to the $\ketA{J'=3/2,p'=+}$ state is nearly degenerate with the D2 transition frequency. Driving such a transition produces near-unity loss to excited rotational states. This loss was observed prior to installing the compensation coils in D2. After installing the compensation coils, we no longer observed this loss as indicated in section \ref{sec:Deceleration}.

Regions S1, D2 (after compensation), and S2 operate at fields under $500$ G. In this low-field regime, the $10\%$ vibrational branching ratio from $\APiHalf (000)$ to the $\XSigma (100)$ limits the optical pumping efficiency \cite{Mengesha2020}. Simulations show that below $\sim 1$T, the population can still be almost entirely transferred between WFS and SFS manifolds using $\lesssim 4$ photon scatters per molecule.  Without vibrational repumping lasers, calculations show $\sim 75\%$ efficiency in transferring SFS to WFS molecules with 4 scattered photons, with the vast majority of the remaining $25\%$ of the population lost to $\tilde{X} (100)$. Less than $5\%$ of the population remains in the SFS manifold after passing through any of these regions. The major difference in efficiency when compared to CaOH is due to the larger vibrational branching to the \ketX{v_1=1} manifold. 

The situation differs qualitatively at magnetic fields above $\sim1$ T, i.e. in regions D1 and D3. In this regime, the Zeeman energy is much greater than the rotational energy spacing, leading to significant rotational mixing between the $\ketA{J' = 1/2, p' = +}$ and $\ketA{J' = 3/2, p' = +}$ manifolds. The lower state in the $\ketA{J' = 1/2, p' = +}$ manifold retains dominant $J' = 1/2$ character, but the upper state in the manifold gains significant $J' = 3/2$ character. This rotational mixing induces a $\sim 10\%$ loss to the \ketX{N''=3} state when pumping through this state. Additionally the large effective g-factor $g'_l$ causes the states that correlate to $\ketA{J' = 1/2, p' = +} $ at zero field to be split substantially. At 2.4~T, they are split by 5.6 GHz and transitions to both components must be driven to drive the entire ground-state WFS manifold toward the SFS manifold. This frequency splitting can be bridged by a high-frequency electro-optic modulator (EOM) and pumping through both excited states requires $\sim 4$ cycles to spin flip all possible ground-state population.

In a high-field regime, not only must we still cycle several photons to drive all possible molecules from WFS to SFS states, but also the loss per cycle is nearly double the low-field value due to the additional loss channel. At 2.4~T, $< 60\%$ of the population can successfully transfer from the WFS to the SFS manifolds at all, due to the substantial rotational and vibrational loss. Calculations show that four scattered photons are sufficient to transfer $\sim 55 \%$ of the population, with only $\sim 5\%$ remaining in the WFS manifold.

\begin{table}[]
\begin{center}
\caption{Comparison of experimentally measured pumping efficiencies to the optimal values calculated from theory for different combinations of pumping light.  The measurements approach the best efficiencies our scheme is capable of as described in Section~\ref{sec:OpticalPumping}. }
\begin{tabular}{ |c|c|c| } 
 \hline
 Light on & Detected Pop. ($\%$) & Optimum (Calc.) ($\%$) \\
 \hline
 S1 & 100 & 100 \\
 S1 + D1& 25(3) & 20 \\ 
 S1 + D1 + D2 & 55(4) & 66 \\ 
 \hline
 Round Trip Pop. & 30(5) & 41\\
 \hline
\end{tabular}
    \label{tab:PumpEff}
\end{center}
\end{table}

Combining the efficiencies for each individual segment of the scheme, and assuming sufficient transit time in all optical pumping regions (i.e. maximally allowed photon cycling), the inefficiencies described in this section imply that $10-15\%$ of the population that enters the ZS slower can be successfully decelerated. In principle, this fraction could be significantly increased by recovering population lost to rovibrational dark states  through vibrational repumping. Though the loss channels are dominated by only two rotational levels ($\ketX{v_1=1,N''=1}$ and $\ketX{v_1=0,N''=3}$), both of these rotational levels will split into separate SFS and WFS manifolds (each with multi-level substructure), making rotational repumping a complex task. The complex Zeeman structure of the $\APiHalf$ state in YbOH significantly impacts both the number of optical cycles necessary to spin-flip population in a high-field ZS decelerator and the amount of loss that those cycles induce. 

While we have focused specifically on YbOH, the structural features that complicate ZS decelerator appear to be generic to many heavy molecules that have been proposed for next-generation precision measurements, including YbF \cite{Fitch_2020,DUNFIELD1995433}, BaF \cite{Aggarwal2018,Steimle2011}, and BaOH \cite{Denis2019,TANDY201144}.

\section{Deceleration of YbOH} 
\label{sec:Deceleration}
Overcoming the complications described above, we demonstrate ZS deceleration of YbOH. To characterize the overall efficiency associated with the deceleration scheme, we confirm the performance of each optical pumping step. We compare the ratio of detected WFS signal in various pumping configurations, as shown in Table~\ref{tab:PumpEff}. For these measurements, the state preparation light S1 is always on. First, we turn on the high-field pumping region D1, which leads to depletion of the WFS population. Then we keep the D1 light on and add the low-field pumping region D2, reviving the population that was depleted by D1. The ratios we measure are similar to what calculations predict for optimum pumping efficiencies. By combining similar measurements with D3, D4, and S2, we find that approximately $\sim 10\%$ of the population succesfully completes two full ZS cycles, in good agreement with the calculations in Section \ref{sec:OpticalPumping}.

\begin{figure}[tb]
    \centering 
    \includegraphics[width=0.6\columnwidth]{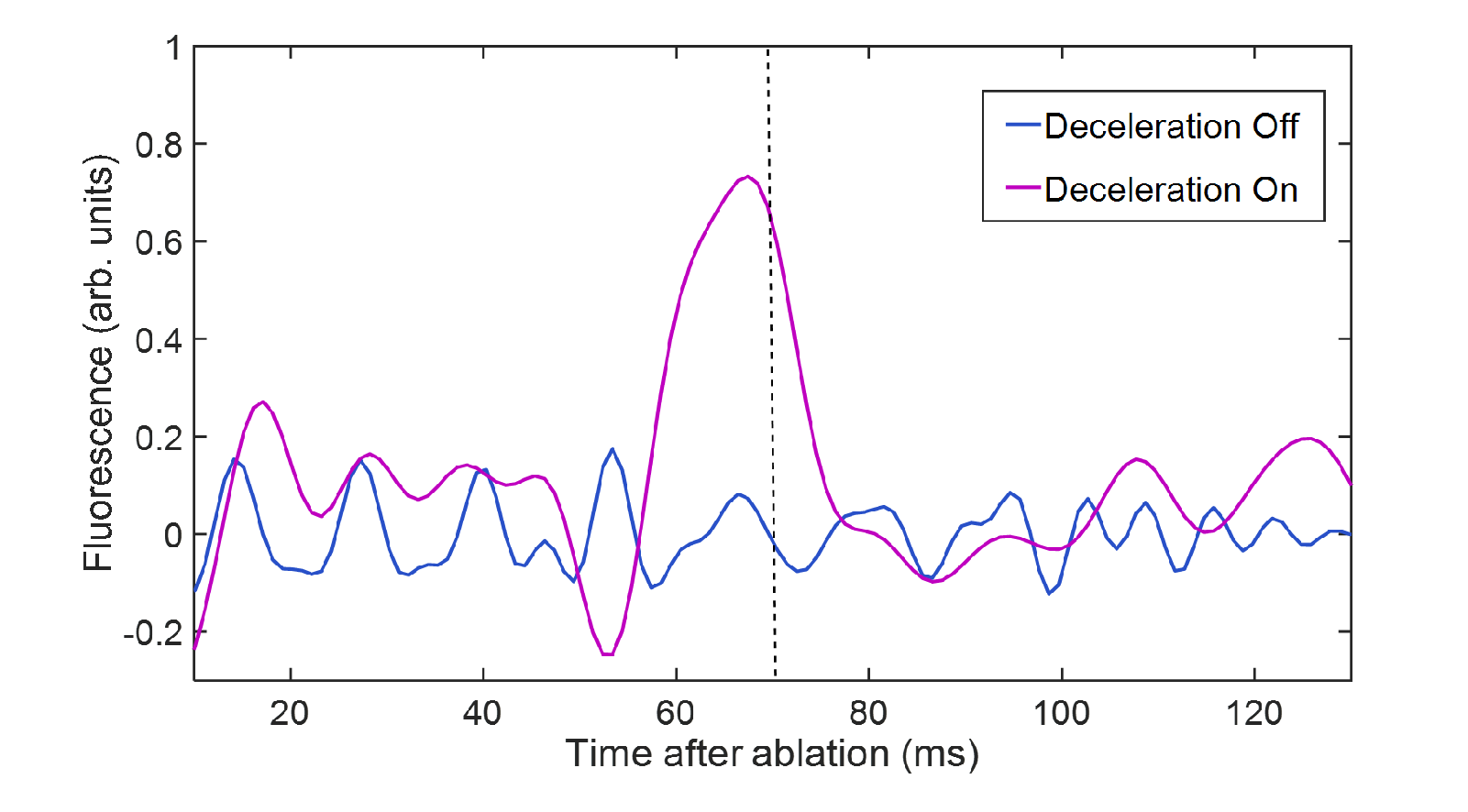}
    \vspace*{-5mm} 
    \caption{ Demonstration of Zeeman-Sisyphus deceleration of YbOH molecules. Both traces are taken in Doppler-sensitive configuration looking at the 18(3) m/s velocity class. The blue trace was taken with only S1 light on, so no deceleration occurs. The purple trace was taken with the full ZS pumping scheme, and reveals decelerated population in this velocity class. The dashed line demonstrates the earliest time of arrival for molecules produced at this speed. 
}
    \label{fig:combinedslowing}
\end{figure}

After optimizing the optical pumping efficiency in each region, we detect individual velocity classes of molecules exiting the decelerator using Doppler-sensitive laser excitation. Figure~\ref{fig:combinedslowing} shows a representative laser-induced fluorescence signal for the velocity class at 18(3)~m/s. Accounting for optical collection efficiency and quantum efficiency of our detector (a photomultiplier tube), We determine that $\sim 100$ molecules per pulse in this single velocity class are decelerated. As expected from the slowing mechanism, the molecules arrive in the detection region earlier than an arrival time that would be expected for molecules that were not slowed, but, rather produced in the beam source at this velocity. This early arrival necessarily means that the detected molecules originate at a higher velocity, providing further confirmation that the detected molecules were decelerated. The molecules detected at 18~m/s originate at velocities near 30 m/s, as expected from the known magnetic potential energies experienced by the molecules. Note that this 12~m/s decrement of velocity would have required $\sim$3,000 photons if radiative slowing were used, while the average molecule undergoing ZS deceleration scattered $\lesssim 10$~photons. This $>100\times$ gain in efficiency per optical photon scatter would be critically important for molecules that do not have high closed optical cycles.

\section{Conclusion}
In this work, we examine the complex Zeeman structure of the heavy molecule YbOH and investigate the effects of this structure on Zeeman-Sisyphus deceleration. We realize deceleration of YbOH molecules, despite the operational complexity that the YbOH level structure creates. The deceleration observed represents the largest velocity change produced in a YbOH beam yet, and could be further improved with a lower beam source velocity (e.g. by lower cryogenic temperatures), more rovibrational rempumping, and/or adding a few more magnetic stages (conveniently at lower field). Modeling shows that closing the rotational and vibrational loss channels with additional frequencies would increase the number of molecules below 20 m/s by an order of magnitude. Laser-induced enhanced production of YbOH\cite{Jadbabaie2020} and an optimized initial beam velocity would increase the number further to $\geq 10^4$ decelerated molecules. With this improved decelerator performance, and typical trap capture efficiencies for polyatomic molecules~\cite{Vilas2022}, it is likely possible to capture $\gtrsim 100$ molecules in an optical dipole trap, a number that is sufficient to obtain an electron EDM sensitivity that surpasses the current limit of $d_e \lesssim 10^{-29}$~e~cm~\cite{ACME2018}.

These results provide insight into the deceleration dynamics of complex molecules. In particular, it is quite common for heavy species to have excited-state $g$-factors sufficiently large to produce Zeeman splittings comparable to $\Lambda$-doubling splitting at modest fields. Molecules in this class will likely require additional repumping lasers to avoid inefficiencies in Zeeman-Sisyphus deceleration schemes that involved high-field optical pumping. This class includes many molecules of broader interest, such as YbF~\cite{Fitch_2020,DUNFIELD1995433}, BaF~\cite{Aggarwal2018,Steimle2011}, BaOH~\cite{Denis2019,TANDY201144}, and WC~\cite{Lee2009,Sickafoose2002,WCZeeman}. Many other experimentally relevant heavy species (e.g. $\rm{YbOCH_3}$~\cite{Augenbraun2021Observation}, $\rm{YbCH_3}$ \cite{Chamorro2022}, YbSH~\cite{Augenbraun2020ATM}, $\rm{BaCH_3}$\cite{Chamorro2022}, RaF \cite{Isaev2010}, RaOH\cite{isaev2016laser}, HgF \cite{Yang2019}, HgOH \cite{Mitra2021}) require further examination of excited state structure to assess the feasibility of Zeeman-Sisyphus deceleration. In contrast, light species of general interest for precision measurements and/or quantum computation (such as MgF~\cite{Xu2016,Chae2021}, MgNC~\cite{Wright1999,Norrgard2019,Norrgard2020MgNC}, $\rm{CaOCH_3}$~\cite{Crozet2002,mitra2020direct}) generally have small $\Lambda$-doubling and will not suffer from the observed Zeeman-induced rotational loss channels, even at higher magnetic field strengths.

To eliminate the observed problematic optical spin-flip transitions, one could replace the high-field optical pumping with microwave transitions. For essentially any molecule with a $^2\Sigma^+$ ground state, this would mean replacing the D1/D3 (see Figure~\ref{fig:ZSscheme}) light with a microwave source to drive population from the WFS to the SFS manifold directly. For decelerators operating below 3~T, the required microwave tone is in a convenient technological range of $f \lesssim 80$~GHz. Optical pumping in the low field regions, where optical spin-flips are more efficient and do not lead to rotational loss, maintains the irreversability of the method. This modification would be species independent, as any $^2\Sigma^+$ ground state will have similar Zeeman tuning and could be decelerated similarly, opening the door to simultaneous Zeeman-microwave deceleration of multiple species.

\section{Acknowledgments}
The authors thank Zack D. Lasner for helpful discussions. The authors gratefully acknowledge the Hutzler group (Caltech) for helpful inputs and for providing pressed powder ablation targets. This work was supported by the Betty Moore Foundation, the Alfred P. Sloan Foundation, and the NSF. HS and TO acknowledge financial support from the Ezoe Memorial Recruit Foundation. BLA acknowledges financial support from the W. M. Keck Foundation and AFOSR.

\typeout{}
\bibliography{ZS_library}

\end{document}